\documentclass[10pt,a4paper]{article} % 10pt is ignored!
\usepackage[dvips]{graphicx}
\usepackage{amsfonts,amsmath,amssymb}
\usepackage{hyperref}
\usepackage{color,soul}  %for highlighting text  \hl{highlighted text}

\begin{document}

\title{The laws of thermodynamics and information for emergent cosmology}
\author{M. Hashemi$^1$,\, S. Jalalzadeh$^1$\footnote{s-jalalzadeh@sbu.ac.ir}\,\,\,\,and\,\,\,S. Vasheghani Farahani$^2$\\
\\\begin{small}$^1$ Department of Physics, Shahid Beheshti University, G. C., Evin,Tehran, 19839, Iran\end{small}
\\\begin{small}$^2$ Department of Physics, Tafresh University, Tafresh, P.O. Box 39518-79611, Iran\end{small}}

\maketitle
%%%%%%%%%%%%%%%%%%%%%%%%%%%%%%%%%%%%%%%%%%%%%%%%%%%%%%%%%%%%%%%%%%%%%%%%%%%%%%%%%%%%%%%%%%%
\begin{abstract}
The aim here is to provide a set of equations for cosmology in terms of information and thermodynamical parameters. The method we implement  in order to describe the universe is a development of  Padmanabhan\rq{}s approach which is based on the fact that emergence of the cosmic space is provided by the evolution of the cosmic time. In this line we obtain the Friedmann equation or its equivalent the conservation law in terms of information by the implementation of Laundauer\rq{}s principle or in other words the information loss/production rate. Hence, a self consistent description of the universe is provided in terms of thermodynamical parameters. This is due to the fact that in this work the role of information which is the most important actor of all times, has stepped in to cosmology. We provide a picture of the emergent cosmology merely based on the information theory. In addition, we introduce a novel entropy on the horizon, which can also generalize Bekenstein-Hawking entropy for the asymptotic holographic principle.\vspace{5mm}\\%\noindent\\
\vspace{0.8mm}\newline
%\textbf{Keywords}: Emergent Cosmology, Thermodynamics Laws, Landauer principle, Holographic principle
\end{abstract}

%%%%%%%%%%%%%%%%%%%%%%%%%%%%%%%%%%%%%%%%%%%%%%%%%%%%%%%%%%%%%%%%%%%%%%%%%%%%%%%%%%%%%%%%%%%%%%%%%%%%
\section{ Introduction }
\par
Thermodynamics is not anymore a stranger in cosmology. As a matter of fact thermodynamics provides the backbone for all analysis in the context of emergent universe models. The basis for this statement was established by Sakharov in 1967 who first pointed out that instead of spacetime, it is better to talk of spacetime atoms, which is only possible by the language of thermodynamics, see Ref \cite{Sakharov}.  This statement paved the way for scientists to work out Einstein\rq{}s field equations based on the thermodynamic equation $\delta Q=TdS$ known as the Clausius relation \cite{Jacobson}. Note that  $T$ is the Unruh temperature which is observed by an accelerating observer inside the horizon, and $\delta Q$ is the energy flux across the horizon \cite{Unruh}. The fact of the matter is that the Einstein\rq{}s field equations are now understood as the spacetime equations of state.

\par Recently due to the efforts of Verlinde a new interpretation of the spacetime has been provided. This is in a sense that by relying on the information extracted from the holographic screen, three of the fundamental equations of physics, (Newton\rq{}s law of gravity, Newton\rq{}s second law, Einstein\rq{}s field equations)  were seen. Verlinde interpreted the gravitational force as an entropic force which is non-fundamental; this is due to the change of information on the holographic screen, see Ref \cite{Verlinde} for details. However, Padmanabhan argued with the definitions provided by Verlinde, by bringing up two issues; the first was on the general covariance, and the second was on the finite systems e.g. the Sun-Earth system\cite{PadmanabhanMPL}. The former issue was raised based on the fact that in general covariance, space and time are treated within the same body, but Verlinde treated them differently. The later issue was raised due to the fact that our every day experience does not prove that finite systems emerge, which contradicts with Verlinde\rq{}s statement that space emerges everywhere, both around finite or infinite systems. However, by selecting the cosmic time in cosmology, the two issues regarding general covariance and finite systems are out of the question. Padmanabhan pointed out that in cosmology by selecting a cosmic time these two issues do not stand. In other words, Padmanabhan stated that in cosmology, the ``Cosmic space
\footnote{The cosmic space volume is an evolutionary volume in respect to the cosmic time. This statement works only if the   CMB is observed homogeneous and isotropic. As long as the CMB is proved to be homogeneous and isotropic for the geodesic  observer, the respected time is the cosmic time. Note that the evolution of the cosmic space regarding the cosmic time may be subject to authors.}
is emergent as cosmic time progresses" which means that the expansion of the universe works out as long as the holographic equipartition stands, see Refs \cite{Padmanabhan1206,Padmanabhan1207} for details. In this line the first proposal for the emergent cosmology was issued by Padbanabhan as \cite{Padmanabhan1206}
\begin{eqnarray}
\frac{dV_{\rm H}}{dt}=L^2_{P}(N_{\texttt{sur}}-N_{\texttt{bulk}})\label{1},
\end{eqnarray}
where $dt$ and $dV_{\rm H}$ are the variations of the cosmic time and cosmic volume respectively. Note that in Padmanabehan's proposal the cosmic volume has been taken as the Hubble horizon volume.
$N_{\texttt{sur}}$ represents the number of surface degrees of freedom which is equal to $A_H/L^2_{P}$, where $A_H$ equals to $4\pi H^{-2}$ denotes the cosmic sphere\rq{}s area, $H$ is the Hubble parameter, $L_{P}=\sqrt{{\hbar G/ c^3}}$ is the Planck length. The parameters $\hbar$, $G$ and $c$ are  the reduced Planck constant, the gravitational constant, and the speed of light in the vacuum respectively. The bulk degrees of freedom which satisfies the equipartition law of energy is obtained as
\begin{eqnarray}\label{2}
N_{\texttt{bulk}}=\frac{2}{k_{\rm{B}}T}E_{\texttt{Komar}}, 
\end{eqnarray}
the temperature corresponding to the hubble horizon is $T_H=H/2\pi$, and $E_\texttt{Komar}$ is the Komar energy contained inside the bulk of a perfect fluid obtained by 
\begin{eqnarray}\label{2.2}
E_{\texttt{Komar}}=2\int_{\upsilon}dV_{\rm {cs}}[T_{\mu\nu}-{1\over2}\mathcal{T}g_{\mu\nu}]u^{\mu}u^{\nu}=|(\rho+3\rm{p})V_{\rm {cs}}|,
\end{eqnarray}
where $T_{\mu\nu}$ is the energy-momentum tensor, $\mathcal{T}$ is the trace of $T_{\mu\nu}$, $u^{\mu}$ is the 4-vector velocity, and $V_{\rm {cs}}$ denotes the cosmic space volume. The reason for choosing $cs$ as the indice is due to the fact that we have various cosmic space volumes, where (for Padmanabhan proposal we have $V_{\rm {cs}}=V_H=4\pi H^{-3}/3$).
 Note that $E_\texttt{Komar}$ is the source for gravitational acceleration\cite{PadmanabhanCQG2004}.
\par
In general, the dynamical emergent equation is
\begin{eqnarray}\label{3}
\frac{dV_{\rm {cs}}}{dt}=f(N_{\texttt{sur}},N_{\texttt{bulk}}),
\end{eqnarray}
where $f(N_{\texttt{sur}},N_{\text{bulk}})$ is an arbitrary function with respect to the asymptotic holographic principle. Equation (\ref{1})
is the simplest accessible form of Eq. (\ref{3}). In this stage it is worth stating other proposals which prove adequate for better understanding the context of the present study. We start with the proposal issued by Sheykhi \cite{Sheykhi} who provided his suggestion based on the apparent horizon as
\begin{eqnarray}
\frac{dV_{\rm A}}{dt}=L^2_{P} \frac{R_{\rm{A}}}{H^{-1}}(N_{\texttt{sur}}-N_{\texttt{bulk}}) ,\label{4} 
\end{eqnarray}
where $R_{\rm{A}}=[H^2+\frac{k}{a^2}]^{-{1 \over 2}}$ is the apparent horizon radius in the FLRW background. Note that 
the apparent horizon is a boundary surface that allows inward light rays enter it, but prevents outward light rays exiting it. 
The constant $k$ could take values equal to $-1$ in case of an open universe, $1$ in case of a closed universe or $0$ in case of a flat universe \cite{Bak:1999hd}.
$V_{\rm {A}}=4\pi R_{\rm{A}}^3/3$ shows the apparent cosmic space volume.
Similar to the Padmabanbhan proposal $N_{\texttt{sur}}$ equals to $A/L_P^2$, where $A$ equals to $4\pi R_A^2$. The temperature corresponding to the apparent horizon is the Kodama-Hayward temperature ($T_{\rm{A}}$) obtained by {\cite{Cai:2005ra}}
\begin{eqnarray}\label{2.1}
T_{\rm{A}}=\frac{|\kappa|}{2\pi}
=\frac{1}{2\pi R_A}\left(1-\frac{{\dot {R}}_{\rm{A}}}{2HR_{\rm{A}}}\right),
\end{eqnarray}
where $\kappa$ is dynamical surface gravity and the dot denotes derivative in respect to time. Therefore, the bulk degrees of freedom is  $N_{\texttt{bulk}}=\frac{2}{k_{\rm{B}}T_A}E_{\texttt{Komar}}$.
Although this equation works fine, but is however not consistent with Padmanabhan\rq{}s proposal. The reason for this is that the RHS of equation (\ref{4}) is not totally based on thermodynamical quantities. Strictly speaking,  a thermodynamical interpretation of the term ${R_{\rm{A}}}/{H^{-1}}$ is not simply achieved. Yang et al. suggested  \cite{Yang}
\begin{eqnarray}
\frac{dV_{\rm H}}{dt}=f(N_{\texttt{sur}}-N_{\texttt{bulk}}, N_{\texttt{sur}}),\label{6} 
\end{eqnarray}
where for details on the function $f(\Delta N, N_{\texttt{sur}})=L^2_{P}\frac{{\Delta N}/{\alpha}+\tilde\alpha K\left({N_{\texttt{sur}}}/{\alpha}\right)^{1+\frac{2}{1-n}}}{1+2\tilde\alpha K\left({N_{\text{sur}}}/{\alpha}\right)^{\frac{2}{1-n}}}$ , see Ref \cite{Yang} and the appendix. The proposition issued by Eune et al. focuses on the corrected form of the horizon volume \cite{Eune}
\begin{eqnarray}
\frac{dV_{\rm A}}{dt}=L^2_{P}f_k(t)(N_\texttt{sur}-N_\texttt{bulk}),\label{7} 
\end{eqnarray}
where $f_{\rm{k}}(t)$ is the deviation volume coefficient from the flat universe, defined as
\begin{eqnarray}\label{7.1} 
f_{\rm{k}}(t)=\frac{V_{\rm{A}}}{V_k}\left[\frac{\frac{\dot{R}_AH^{-1}}{R_A}-\frac{R_A}{H^{-1}}\frac{V_k}{V_{\rm{A}}}+1}
{\frac{\dot{R}_AH^{-1}}{R_A}+\frac{R_A}{H^{-1}}\frac{V_k}{V_{\rm{A}}}-1}\right],
\end{eqnarray}
where we have $V_{\rm{k}}=2\pi a^2[\sqrt{k}a\arcsin{(\sqrt{k}R_{\rm{A}}/a)}-kR_{\rm{A}}^2H]$. Note that $k$ is as stated earlier in the text. Moreover, a thermodynamic interpretation for $f_k(t)$ is not simple.
\par
In a work prior to the present study we have proposed \cite{HJ}
\begin{eqnarray}
\frac{dV_{\rm{A}}}{dt}={2 L^2_{P}}\frac{T_{\rm{A}}}{T_{\rm{H}}}(N_{\texttt{sur}}-N_{\texttt{bulk}}),\label{5}
\end{eqnarray}
where $T_{\rm{H}}=(2\pi HR_{\rm{A}}^2)^{-1}$ is the cosmological horizon temperature with non-dynamical radius which is measured by a comoving observer \cite{Hu} which by implying an apparent horizon (which is considered as the most appropriate boundary in application to thermodynamics) alongside Kodama-Hayward temperature  (which is a temperature for an evolving horizon implied in cosmology) will enable us to write the RHS of the dynamical emergent equation based only on thermodynamical parameters.
Note that $T_{\rm A}$ is the physical and working temperature in Eq. (\ref{5}).
As of the spirit of the emergent cosmology where the universe has a tendency towards the holographic principle state, $T_{\rm A}$ tends to the asymptotic temperature $T_{\rm H}$. Due to the presence of $H$ and $R_{\rm A}$ in  Eqs. (\ref{4}) and (\ref{7.1}) 
some difficulties may be observed for providing thermodynamical interpretations. This is due to the fact that Eqs. (\ref{4}) and (\ref{7.1}) do not seem to be completely based on information parameters. Equations  (\ref{1}) and (\ref{8})
  although do describe the evolution of an emergent cosmic space, but still lack generality. This generality is observed in Eq.
  (\ref{5}) justifying its importance.
For more proposals please see\cite{otherproposals}).
\par 
In the 70s, studying black hole physics became very hot due to the discovery of a mysterious relation between gravity and thermodynamics \cite{Padmanabhan:2010xh, Padmanabhan:2010rp}. After facing the information paradox while measuring the transmission information through the horizon, measuring the transmitted information from the horizon became a big issue. In the emergent theory due to the assumption of the existence of space atoms, we are hopeful that not only a better description for the universe could be provided, but also the quantum gravity theory could be established. The variation of the cosmic space volume, the space atoms go in and out the horizon. These movements create or destroy information inside the bulk. Space atoms which go in and out the horizon are not recoverable due to the horizon's notion. Now since these inward and outward space atoms are non-recoverable, the change in information that they provide, would generate entropy \cite{Padmanabhan:2010xh,Padmanabhan:2009vy,Dowker:2014xga, Padmanabhan:2008zza}.
\par
In the present work, we discuss the information interpretation of the second law of thermodynamics based on Landauer principle. In fact, the general idea of this work is based on this principle. This principle states that information is a physical concept. As a matter of fact this principle came to save the second law of thermodynamics from criticizers, where the most important one of all was Maxwell\rq{}s demon, see e.g. \cite{Szilard,C�pek}.  However, it was Landauer who saved the day for the second law of thermodynamics by stating that if information on a system is so how deleted that it is impossible to undelete it, entropy is created. Note that Landauer was highly inspired by Szilard\rq{}s engine, who had proved that information is not an intrinsic concept, but has relations with the outside world, see \cite{Szilard}. This entropy which is created by the loss of information is obtained by
\begin{eqnarray}\label{8}
\Delta S=-k_{\rm{B}}\ln2\  \Delta I,
\end{eqnarray}
where $\Delta I$ denotes of information loss. 
This means that for every one bit of destroyed data, the entropy increases as much as $k_{\rm{B}}\ln2$, see \cite{Landauer}. For an intensive reading on Landauer principle, see e.g. \cite{sagawa, DuncanFP2007, DuncanEntropy2004}.

%%%%%%%%%%%%%%%%%%%%%%%%%%%%%%%%%%%%%%%%

\section{The information interpretation of the conservation law}
\par The standard model for cosmology is based on two main equations \footnote{For simplicity, the natural units $k_{{B}}=c=\hbar=1$ are used throughout this section.}
\begin{eqnarray}
{1 \over {R_{\rm{A}}^2}}=H^{2}+\frac{k}{a^2}&=&\frac{8\pi}{3}L_P^2\rho,\\
\dot{H}+H^2&=&-\frac{4\pi}{3}L_P^2(\rho+3\rm{p}),
\end{eqnarray}
where, the first is known as Friedmann\rq{}s equation and the second is known as Raychaudhuri\rq{}s equation. However, Friedmann\rq{}s equation could be derived when Raychaudhuri\rq{}s equation is combined with the conservation law. Note that the conservation law for a universe with a perfect fluid matter content is expressed as 
\begin{eqnarray}\label{10}
\begin{cases}
\dot \rho+3H(\rho +\rm{p})=0 &\quad \texttt{in terms of }H,\\ \\
\dot \rho+2\frac{\dot R_{\rm{A}}}{R_{\rm{A}}}\rho=0 &\quad \texttt{in terms of}\,\, R_{\rm{A}}.
\end{cases}
\end{eqnarray}
In this stage the intention is to write an expression for Eq. (\ref{10}) in terms of the first law of thermodynamics  
\begin{eqnarray}\label{10.1}
dE+\rm{p}dV=TdS
\end{eqnarray}
Before proceeding it is worth providing  an example; consider a comoving volume defined as, $V_c=\frac{4\pi}{3}a^3$, therefore the conservation law would take the form 
\begin{eqnarray}\label{11} 
\dot \rho+\frac{\dot V_c}{V_c}(\rho +\rm{p})=0,
\end{eqnarray}
now if this volume has the Misner-Sharp energy \cite{MS}, we have  
\begin{eqnarray}
dE+\rm{p}dV_c=0.\label{13}
\end{eqnarray}
Equation (\ref{13}) is as the form of the first law of thermodynamics. It could also be deduced from Eq.  (\ref{13}) that $\rm{dS}$ is equal to zero, meaning that the entropy is constant. In other words, a universe defined by the comoving volume with energy $\rm{E}$ has no entropy production.
\par
Now it is time to implement the findings of the present study to a more realistic universe. All of the laws of thermodynamics work perfect for the apparent horizon, the desired boundary here is chosen accordingly (as the apparent horizon). Note that the apparent horizon  has been defined below Eq. (\ref{4}). In this line we start by dividing both sides of the first law of thermodynamics (Eq. \ref{10.1}) by the apparent horizon volume ($V_{\rm{A}}=\frac{4\pi}{3}R_A^{3}$), which gives
\begin{eqnarray}\label{15} 
\dot \rho+3(\rho +\rm{p})\frac{\dot R_{\rm{A}}}{R_{\rm{A}}}=\frac{T_{\rm{A}}\dot S}{V_{\rm{A}}}.
\end{eqnarray}
It is clear that the entropy is not constant, which is due to the selection of the apparent volume. Now the conclusion from all of this is that the apparent horizon volume with a fixed entropy is out of the question in the present study. What we need here is something that exhibits entropy production. The fact of the matter is that  in contrast to a comoving volume there is an entropy production/loss for an observer who travels with the apparent horizon. The question that arises here is how the new entropy would look. To answer this question we must go down to facts; we all know that for the de Sitter universe holographic principle holds ($N_\texttt{sur}=N_\texttt{bulk}$) \cite{Padmanabhan:2013nxa}, hence due to the dynamical equation for emergence we can take $dV_A/dt$ equal to zero. This results in having a constant radius, which leads to conclude that the entropy is constant
This could be readily noticed from Eqs. (\ref{10}) and (\ref{15}) which leads to $\dot S \propto\dot R$.
. It is known that  the de Sitter universe entropy $S_{dS}$  is equal to $3\pi/\Lambda$ which is constant, see e.g. ref {\cite{Bousso}}. Another fact that supports our proposal for the entropy rate is that for vacuum, the information production/loss rate must be zero. Note that for vacuum (cosmological constant $\Lambda$) the equation of state is $\rho+\rm{p}=0$. The last fact is that for the Minkowski spacetime, due to the flat geometry, information is neither produced nor lost. These facts in addition to the fact that entropy is non-constant justifies our proposal for the information production/loss, defined as  
\begin{eqnarray}\label{16}
\frac{dI}{dt}= \frac{3HV_{\rm{A}}-f(N_\texttt{sur},N_\texttt{bulk})}{T_{\rm{A}}\,\ln2}(\rho+\rm{p}),
\end{eqnarray}
where $f(N_\texttt{sur},N_\texttt{bulk})$ has been defined in Eq. (\ref{3}). It is worth reminding that when $3HV_A$ is equal to $f(N_\texttt{sur},N_\texttt{bulk})$, the numerator on the RHS of Eq. (\ref{16}) would be zero, resulting in the Milne universe which is part of the Minkowski spacetime. Now the rate of change for the entropy inside the apparent horizon due to the Landauer Principle (\ref{8}) is
\begin{eqnarray}\label{17}
\frac{dS}{dt}=-\ln2\frac{dI}{dt}= \frac{f-3HV_{\rm{A}}}{T_{\rm{A}}}(\rho+\rm{p}).
\end{eqnarray}
One may reproduce the conservation law (\ref{10}) based on the first law of thermodynamics (\ref{13}), together with the evolutionary parameters expressed by Eqs. (\ref{3}) and (\ref{16}).
\par
To comply with the aims of the present study which is to write the Friedmann\rq{}s equation for an apparent horizon, we write all emergent equations in terms of information. The two fundamental equations for describing the cosmic space in the emergent perspective are
\begin{eqnarray}
\frac{dV_{\rm{A}}}{dt}&=&f(N_\texttt{sur},N_\texttt{bulk}),\label{20}\\\nonumber\\
\frac{dI}{dt}&=& \frac{3HV_{\rm{A}}-f}{T_{\rm{A}}\,\ln2}(\rho+\rm{p}).\label{21}
\end{eqnarray}
The dynamical equation of emergent gravity (\ref{20}) is equivalent to the Raychaudhuri equation in the standard model of cosmology, and the rate of information change of the apparent horizon (\ref{21}) is equivalent to equation (\ref{10}). Therefore, one can say that Eqs. (\ref{20}) and (\ref{21}), would lead to the equations for the standard model of cosmology.
\par
In order to establish the idea that Equations (\ref{20}) and (\ref{21}) define emergent cosmology, one more thing has to be done. That is on the parameter $H$ In Eq. (\ref{21}) which its information aspect has not been well defined. To proceed with this we have to deal with the last degree of freedom for the system,$f$, located in Eq. (\ref{3}). Therefore the proposal expressed by Eq. (\ref{5}) is implemented. This provides an information nature for all parameters of Eq. (\ref{21}). It is worth noting that proposal (\ref{5}) is well chosen due to the fact that it is more in line with the nature of the emergent cosmology than the others. One may follow the other proposals in the appendix.
In this line we should find the rate of change for the entropy by combining the Landauer Principle (\ref{17}) and the holographic Raychaudhuri equation (\ref{5}), which is
\begin{eqnarray}\label{31}
\frac{dS}{dt}=\left[\frac{2}{T_{\rm{H}}}L^2_{P}(N_{\texttt{sur}}-N_{\texttt{bulk}})-\frac{3HV_{\rm{A}}}{T_{\rm{A}}}\right](\rho+\rm{p}).
\end{eqnarray}
\par
It is instructive to check the generality of Landauer\rq{}s entropy (Eq. (\ref{17})). Equation (\ref{10}) combined with  Eq. (\ref{15}) gives
\begin{eqnarray}\label{36}
\dot S=\frac{V_{\rm{A}}}{T_{\rm{A}}}\left[\dot \rho+3(\rho+\rm{p})\frac{\dot R_{\rm{A}}}{R_{\rm{A}}}\right]=\frac{(\rho+3\rm{p})V_{\rm{A}}}{T_{\rm{A}}}\frac{\dot R_{\rm{A}}}{R_{\rm{A}}},
\end{eqnarray}
where the dot denotes derivative in respect to time. Equation (\ref{36}) could be written in terms of the surface and bulk degrees of freedom 
\begin{eqnarray}\label{37}
\begin{array}{cc}
\dot S=\frac{1}{4}\frac{N_\texttt{bulk}}{N_\texttt{sur}}\dot N_\texttt{sur},
\end{array}
\end{eqnarray}
Note that $N_\texttt{bulk}$ and $N_\texttt{sur}$ have been defined in Eq. (\ref{2}) and its preceding paragraph. Equation (\ref{37}) could also be written as the change of entropy in terms of the surface degrees of freedom
\begin{eqnarray}\label{38}
dS=\frac{1}{4}\frac{N_\texttt{bulk}}{N_\texttt{sur}}dN_\texttt{sur}.
\end{eqnarray}
It could readily be noticed that when the holographic condition ($N_\texttt{sur}=N_\texttt{bulk}$) applies, the well-known Bekenstein-Hawking entropy ($S=\frac{1}{4}\frac{A}{L^2_{P}}$) is resulted from Eq. (\ref{38}). Hence interestingly, Eq. (\ref{38}) can be considered as a generalization of the Bekenstein-Hawking entropy which holds not only for the holographic principle but also for the asymptotic holographic principle.
\par
In this stage it is worth writing the emergent equations in the form proposed in the present study as
\begin{eqnarray}\label{40}
\begin{cases}
\frac{dV}{dt}={2 L^2_{P}}\frac{T_{\rm{A}}}{T_{\rm{H}}}(N_{\texttt{sur}}-N_{\texttt{bulk}}),
\\\\ dI=\frac{-1}{4\,\ln2}\frac{N_\texttt{bulk}}{N_\texttt{sur}}dN_\texttt{sur},
\end{cases}
\end{eqnarray}
which is accompanied by the first law of thermodynamics (\ref{10.1}).
\par
To comply with the aims of the present study which is to obtain an expression for the cosmological equations, we start from the holographic Raychaudhuri equation (\ref{5}). But first we should recall two parameters which are the Kodama-Hayward temperature \cite{Cai:2005ra,Hayw1,Hayw2,Helou:2015yqa} and the Komar energy of the universe. As stated earlier, in order to measure the temperature of the evolving horizons it is suitable to use the Kodama-Hayward temperature, $T_{\rm{A}}=\frac{1}{2\pi R_{\rm{A}}}\left(1-\frac{\dot R_{\rm{A}}}{2HR_{\rm{A}}}\right)$ , which is measured by the Kodama observer. It is worth noting that in emergent cosmology due to the asymptotic de Sitter behavior the existence of dark energy is compulsory, see \cite{Padmanabhan1207}. Thus, the temperature for accelerated universe with perfect fluid will never be zero. This means that the total energy density, $\rho$,  would never be equal to triple of the total pressure, $3\rm{p}$. Hence the temperature would never become zero.
The Komar mass of the accelerated universe expressed as \cite{Padmanabhan1207,PadmanabhanCQG2004}
\begin{eqnarray}
E_{\texttt{Komar}}=|(\rho+3\rm{p})V_A|=-(\rho+3\rm{p})V_{\rm{A}}. \label{25}
\end{eqnarray}
Substitute the definitions for the surface and the bulk degrees of freedom (\ref{2}) in the holographic Raychaudhuri equation (\ref{5}). The result would be 
\begin{eqnarray}\label{29}
4\pi R^2_A\dot{R}_A=2L^2_{Pl}HR_A\left(1-\frac{\dot{R}_A}{2HR_A}\right)
\left[\frac{4\pi R_A^2}{L^2_{Pl}}+\frac{32\pi^2R_A^5H(\rho+3\rm{p}) }{3(2HR_A-{\dot {R}}_A)}\right],
\end{eqnarray}
where by rearranging leads to the Raychadhuri equation
\begin{eqnarray}\label{30}
\dot{H}+H^2=-\frac{4\pi}{3}L_{P}^2(\rho+3\rm{p}).
\end{eqnarray}
Equation (\ref{30}) is one of the cosmological equations for the standard model of cosmology that we where after, the other one is the  Friedmann equation.
By substituting Eq. (\ref{31}) into the first law of thermodynamics, the conversation law (Eq. (\ref{10})) can be obtained. Clearly, Eqs. (\ref{10}) and (\ref{30}) are sufficient to derive the Friedmann equation, expressed by 
\begin{eqnarray}\label{34}
\frac{1}{R_{\rm{A}}^2} =\frac{8\pi L_p^2}{3}\rho.
\end{eqnarray}
The picture that we have drawn of the cosmological universe by Eqs. (\ref{30}) and (\ref{34}) due to the Landauer Principle and the asymptotic holographic principle, shed light on the universe based on information.
%%%%%%%%%%%%%%%%%%%%%%%%%%%%%%%%%%%%%%%%%%%%%%%%%%%%%%%%%%%%%%%%%%%%%%%%%%%%%%%%
\section{ Concluding remarks}
\par
The main achievement of the present work is writing the equations of cosmology in terms of the information of the system. The starting point of this study was Padmanabhan\rq{}s  proposal on the emergence of cosmic space. He stated that the difference between the degrees of freedom on the surface and the bulk of the horizon, causes the emergence of the cosmic space. He supported his statement by deriving the Raychaudhuri equation for flat FLRW.
The emergent dynamical expression (Eq. (\ref{3})) is the equivalent of the Raychaudhuri equation in the standard model of cosmology. The emergent equation shows the dependence of the cosmic volume space on the cosmic time in terms of the number of degrees of freedom, which itself is based on the information extracted from the system.  The question to be answered is whether if it is possible to write the Friedmann equation or in other words the continuity equation in terms of the information of the system. The key to this question lies in the hands of the Landauer\rq{}s principle and the loss/production of information, where by implementing these, we have obtained the continuity equation. This enabled us to obtain two consistent equations in emergent cosmology based on the information extracted from the surface and bulk of the apparent horizon, see Eq. \ref{40}). 

\par
We obtain a generalized form for the Bekenstein-Hawking entropy both for the holographic principle and the asymptotic holographic principle (Eq. (\ref{38})) enables a more realistic understanding of the horizon\rq{}s entropy.

%%%%%%%%%%%%%%%%%%%%%%%%%%%%%%%%%%%%%%%%%%%%%%%%%%%%%%%%%%%%%%%%%%%%%%%%%%%%%%%%

\appendix 
\setcounter{equation}{0}  % reset counter 
\section*{APPENDIX}  % use *-form to suppress numbering
\renewcommand{\theequation}{A.\,\arabic{equation}}  % redefine the command that creates the equation no.
\par
It is worth investigating other proposals fron $f$ extracted from Eq. (\ref{3}) in terms of the Laundauer entropy (\ref{17}). In this line after checking Laundauer entropy for other proposals it will be seen that the chosen proposal of the present study (\ref{5}) is indeed reasonable. We hope that the last degree of freedom in the dynamical emergent equation (choosing $f$ in Eq. (\ref{3})) eliminates $H$ from Eq. (\ref{17}). Note that the existence of $H$ in Eq. (\ref{17}) puts us in a weak position for issuing an informational interpretation for the emergent cosmological equations (\ref{20}) and (\ref{21}).

\par
Equation (\ref{1}) was proposed  by Padmanabhan. He assumed the hubble horizon ($R_{\rm{H}}=H^{-1}$) to be the boundary of the universe. Therefore, the area and volume of the universe would be expressed as
\begin{eqnarray}\label{A1}
A_H=4\pi H^{-2},\quad V_{\rm{H}}={4\pi \over 3} H^{-3}.
\end{eqnarray}
The temperature corresponding to the horizon is assumed $T_H=H/2\pi$. By calculating $N_{\texttt{sur}}$ and $N_{\texttt{bulk}}$ as defined in the Eq. (\ref{2}) and its preceding paragraph, we have
\begin{eqnarray}\label{A2}
N_{\texttt{sur}}={4\pi \over {L^2_P H^2}}, \quad
N_{\texttt{bulk}}=-{16\pi^2 \over {3 H^4}} (\rho+3\rm{p}).
\end{eqnarray}
Now by substituting these parameters into the Laundauer entropy (\ref{17}) we have
\begin{eqnarray}\label{A3}
\dot S={32\pi^3 \over {3 H^5}} L^2_P(\rho+3\rm{p}) \, (\rho+\rm{p}).
\end{eqnarray}
Unluckily $H$ remains in the Equation disabling a robust thermodynamical interpretation of Laundauer-Padmanabhan Entropy.

\par
Shyekhi's Proposal is Eq.(\ref{4}). He generalized the boundary of the universe to the apparent horizon and provided a general expression for the curvature of the universe. The area and volume for the universe is as of Shyekhi's Proposal
\begin{eqnarray}\label{B1}
A_A=4\pi R_{\rm{A}}^2,\quad V_{\rm{A}}={4\pi \over 3} R_{\rm{A}}^3.
\end{eqnarray}
The Kodama-Hayward temperature of the apparent horizon is assumed to be in the form $T=1/2\pi R_{\rm{A}}$. Similar to Laundauer-Padmanabhan entropy we should calculate $N_{\texttt{sur}}$ and $N_{\texttt{bulk}}$ as defined in Eq. (\ref{2}) and its preceding paragraph
\begin{eqnarray}\label{B2}
N_{\texttt{sur}}={4\pi R_{\rm{A}}^2\over {L^2_P}}, \quad
N_{\texttt{bulk}}=-{16\pi^2 \over 3}R_{\rm{A}}^4 (\rho+3\rm{p}),
\end{eqnarray}
resulting in
\begin{eqnarray}\label{B3}
\dot S={f-3HV_{\rm{A}} \over T_{\rm{A}}} \, (\rho+\rm{p})={32\pi^3 \over 3} R_{\rm{A}}^6 H L^2_P(\rho+3\rm{p}) \, (\rho+\rm{p}).
\end{eqnarray}
It is not surprising that $H$ still exists. We know that by substituting $R_{\rm{A}}=H^{-1}$ in Eq. (\ref{B3}) one could reach Eq.(\ref{A3}). The informational interpretation of this entropy is unclear.

\par
Another extension of Padmanabhan's approach hs been provided by Yang et al. \cite{Yang}. Their proposal (Eq. (\ref{6})) calculates the  Raychaudhuri\rq{}s equation for an arbitrary dimension. For simplicity we focus on the 3+1 dimension universe. Their assumptions for the horizon radius and its temperature is similar to Padmanabhan's approach, see equations (\ref{A1} and \ref{A2}). In 3+1 dimensions, the auxiliary parameters $\alpha$ and $K$  \cite{Yang} are expressed  by  
\begin{eqnarray}\label{C1}
\alpha=1, \quad K={3\Omega_3 \over L^2_P}={4\pi \over L^2_P}.
\end{eqnarray}
Another auxilary parameter $\tilde \alpha$ is the new degree of freedom in the Yang's proposal which meets Padmanabhan's proposal if$\tilde \alpha$  equals zero. Note that for consistency (see Eq. (\ref{3})) we have the constant $L_p^2$ is taken to the RHS of EQ. (\ref{6})\\
 \begin{eqnarray}\label{C2}
f&=&L^2_{P}\frac{{\Delta N}/{\alpha}+\tilde\alpha K\left({N_{\texttt{sur}}}/{\alpha}\right)^{1+\frac{2}{1-n}}}{1+2\tilde\alpha K\left({N_{\text{sur}}}/{\alpha}\right)^{\frac{2}{1-n}}}\\\nonumber
\\\nonumber
&=&\frac{L^2_{P}{\Delta N}+4\pi \tilde\alpha} {1+{8\pi\tilde\alpha \over L^2_p N_{\texttt{sur}}}}
=\frac{f_{\texttt{Padmanabhan}}+4\pi \tilde\alpha} {1+{8\pi\tilde\alpha \over L^2_p N_{\texttt{sur}}}},
\end{eqnarray}
where $f_{\texttt{Padmanabhan}}$ is Padmanbhan's proposal (\ref{1}). Yang's proposal is reduced to the Padmanabhan's suggestion if $\tilde \alpha=0$. By substituting equations (\ref{A1}), (\ref{A2}) and (\ref{C1}) into the Eq. (\ref{C2}) we have

\begin{eqnarray}\label{C3}
\dot S={2\pi  \over { H(1+2\tilde \alpha H^2)}}(\rho+\rm{p}) \left[{16\pi^2 \over {3 H^4}} L^2_P(\rho+3\rm{p})+4\pi \tilde \alpha  \right].
\end{eqnarray}
Equation {\ref{C3}} is not directly based on only thermodynamic parameters.
\par
Eune et al. propsed another propsal by calculating the volume differently. In brief their proposal is described as
\begin{eqnarray}\label{D1}
f=f_k(t){H^{-1}\over R_{\rm{A}}}\,f_{\texttt{Sheykhi}},
\end{eqnarray}
where $f_k$ is defined by Eq. (\ref{7.1}). Except the definition of volume in their approach (and their proposal) other assumptions are the same as of sheykhi's proposal. The entropy would be
\begin{eqnarray}\label{D2}
\dot S=\left({2\pi  \over H} 
\left\{\frac{V_{\rm{A}}}{V_k}\left[\frac{\frac{\dot{R}_AH^{-1}}{R_A}-\frac{R_A}{H^{-1}}\frac{V_k}{V_{\rm{A}}}+1}
{\frac{\dot{R}_AH^{-1}}{R_A}+\frac{R_A}{H^{-1}}\frac{V_k}{V_{\rm{A}}}-1}\right]\right\}
f_{\texttt{Sheykhi}}
-6\pi H R_{\rm{A}}V_{\rm{k}}\right) (\rho+\rm{p}).
\end{eqnarray}
Note that in this appendix the compatibility check between the Laundauer entropy proposed in this work (Eq. (\ref{17})) with other dynamical emergent equations (Eq. \ref{3}) has been carried out. However, other potential proposals for Laundauer entropy may provide an expression based on only thermodynamic parameters.

%%%%%%%%%%%%%%%%%%%%%%%%%%%%%%%%%%%%%%%%%%%%%

\end{document}